\documentclass[]{spie}  

\usepackage{amsmath,amsfonts,amssymb}
\usepackage{graphicx}
\usepackage[colorlinks=true, allcolors=blue]{hyperref}
\usepackage{float}
\usepackage{caption}

\title{A two-band approach to n$\lambda$ phase error corrections with LBTI's PHASECam}

\author[a]{E.R. Maier}
\author[a]{P.M. Hinz}
\author[b]{D. Defr\`{e}re}
\author[a]{S. Ertel}
\author[a]{E. Downey}

\affil[a]{Steward Observatory, University of Arizona, 933 North Cherry Ave, Tucson, AZ 85721, USA}
\affil[b]{STAR Institute, Universit\'{e} de Li\`{e}ge, 17 Alle\'{e} du Six Ao\^{u}t, B-4000 Sart Tilman, Belgium}

\authorinfo{Further author information: Send correspondence to E.M.\\E-mail: erinrmaier@email.arizona.edu}

\begin{document} 
\maketitle
\begin{abstract}
PHASECam is the Large Binocular Telescope Interferometer's (LBTI) phase sensor, a near-infrared camera which is used to measure tip/tilt and phase variations between the two AO-corrected apertures of the Large Binocular Telescope (LBT). Tip/tilt and phase sensing are currently performed in the H (1.65 $\mu$m) and K (2.2 $\mu$m) bands at 1 kHz, and the K band phase telemetry is used to send tip/tilt and Optical Path Difference (OPD) corrections to the system. However, phase variations outside the range [-$\pi$, $\pi$] are not sensed, and thus are not fully corrected during closed-loop operation. PHASECam's phase unwrapping algorithm, which attempts to mitigate this issue, still occasionally fails in the case of fast, large phase variations. This can cause a fringe jump, in which case the unwrapped phase will be incorrect by a wavelength or more. This can currently be manually corrected by the observer, but this is inefficient. A more reliable and automated solution is desired, especially as the LBTI begins to commission further modes which require robust, active phase control, including controlled multi-axial (Fizeau) interferometry and dual-aperture non-redundant aperture masking interferometry. We present a multi-wavelength method of fringe jump capture and correction which involves direct comparison between the K band and currently unused H band phase telemetry. 
\end{abstract}


\section{INTRODUCTION}
\label{sec:intro}  

A significant obstacle faced by ground-based optical and infrared interferometry is rapid Optical Path Difference (OPD) variations introduced by turbulence in the atmosphere\cite{Bouquin08}. This causes variations in the fringe phase, also known as the phase delay, leading to blurring of the fringes and preventing meaningful interferometric measurements. In order to obtain sensitive, accurate interferometric observations, the fringes must be stabilized by reducing the differential OPD between apertures to a fraction of the observing wavelength\cite{Petrov16}. In practice, this is often achieved using a fringe tracker, a device which measures the fringe phase and uses this to correct the differential OPD in real time. However, when these measurements are carried out using quasi-monochromatic fringes, there exists a 2$\pi$ ambiguity in the phase measurement, which leads to a $\lambda$ ambiguity in the OPD correction. This can lead to undetected fringe jumps, which will prevent coherent imaging at the science wavelength. This ambiguity can be avoided with multi-wavelength phase sensing, which will remove the ambiguity out to the first common multiple of all observed wavelengths\cite{Petrov16}.

The Large Binocular Telescope Interferometer (LBTI) is a NASA-funded nulling and imaging instrument designed to coherently combine the two primary mirrors of the Large Binocular Telescope (LBT)\cite{Hill14, Veillet14, Veillet16} for high-sensitivity, high-contrast, and high-resolution infrared imaging (1.5 - 13 $\mu$m)\cite{Hinz16, Hinz18}. The LBTI is equipped with two science cameras: LMIRCam\cite{Wilson08, Skrutskie10, Leisenring12} (the L and M Infrared Camera, 3-5 $\mu$m) and NOMIC\cite{Hoffmann14} (Nulling Optimized Mid-Infrared Camera, 8-14 $\mu$m). The LBTI's fringe tracker is PHASECam\cite{Defrere14}, a near-infrared (1.5 - 2.5 $\mu$m) camera which measures and corrects differential tip/tilt and OPD variations between the two AO-corrected apertures of the LBT. PHASECam carries out phase measurements in both the H (1.65 $\mu$m) and K (2.2 $\mu$m) bands at 1 kHz, but currently only uses the K band phase information for active phase control.

Though PHASECam does see fringe jumps, until now, this quasi-monochromatic approach to phase sensing and control has sufficed as PHASECam has primarily been required for the nulling interferometric observations of the Hunt for Observable Signatures of Terrestrial Planets (HOSTS) survey, a NASA-funded N-band (10 $\mu$m) survey of exozodiacal dust around nearby stars\cite{Defrere16a, Ertel18a, Ertel18b}. Fringe jumps which occur during these observations are easily seen in the null telemetry, and can be corrected manually by the observer. However, manual correction is inefficient. Furthermore, LBTI also utilizes several other observing modes which require active phase control, but which do not possess a mechanism such as the null telemetry for easy detection of fringe jumps. These include imaging, or ``Fizeau'', interferometry, as well as Non-Redundant Aperture Masking interferometry (NRM). 

\begin{figure}[ht!]
\begin{minipage}[c]{.5\textwidth}
\centering
\includegraphics[width=.8\textwidth, trim=0cm .5cm 0cm .5cm, clip=true]{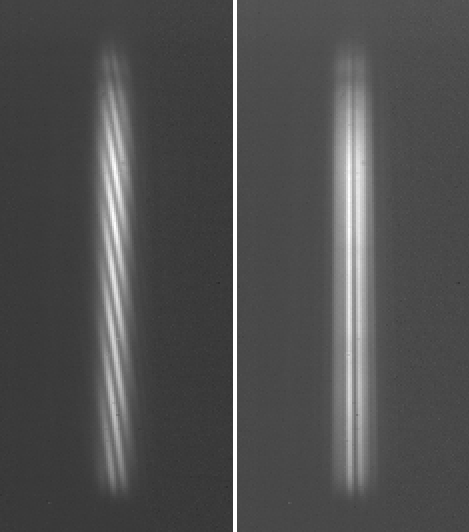}
\vspace*{3mm}
\caption*{(a)}
\end{minipage}
\hfill
\begin{minipage}[c]{0.5\textwidth}
    \includegraphics[width=\textwidth]{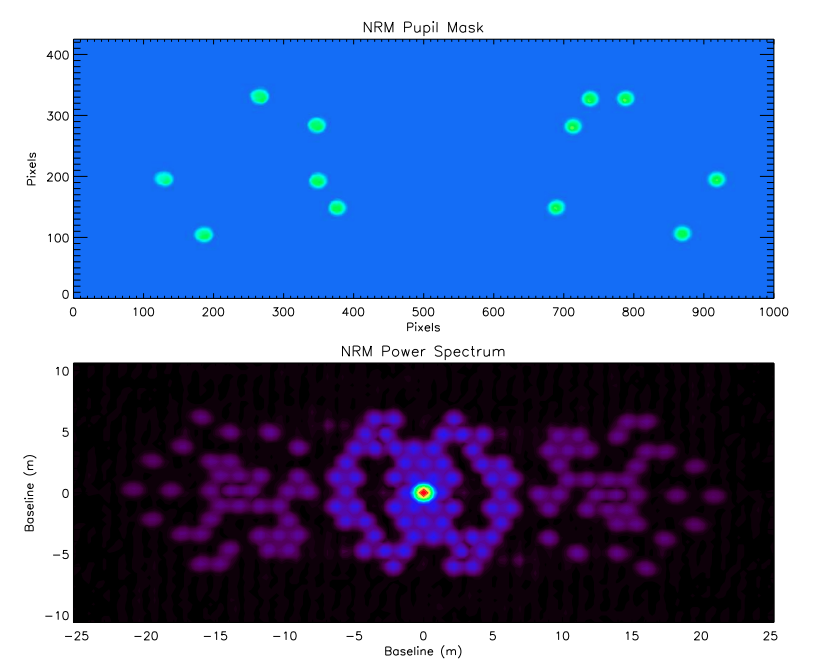}
\vspace*{1.5mm}
\caption*{(b)}
\end{minipage}
\vspace*{7mm}
\caption[angle=90]{(a) \emph{Left}: The ``barber-pole'' Fizeau fringes seen with a 2.8-4.2 $\mu$m L-band grism\cite{Kuzmenko12} after initial detection of the coherence envelope on LMIRCam. \emph{Right}: The approximately vertical Fizeau fringes after minimization of the OPD. See [23] in these proceedings. (b) \emph{Top}: The 12-hole NRM pupil mask observed by LMIRCam using a pupil-imaging lens. \emph{Bottom}: Observed power spectrum of a bright calibrator during LMIRCam observations with the 12-hole mask. The bright central cluster corresponds to intra-aperture baselines up to the single-aperture diameter of 8 m, while the left and right clusters correspond to inter-aperture baselines up to the edge-to-edge mirror separation of 22.8 m. See [10].}
\label{phase-lock-modes}
\end{figure}

Fizeau interferometry utilizes focal plane (multi-axial) beam combination across the entire 22.8 m edge-to-edge LBT mirror separation\cite{Patru17a, Patru17b}. This is in contrast to nulling interferometry, which utilizes pupil plane (co-axial) beam combination across the 14.4 m center-to-center mirror separation. Previously, LBTI had only imaged a small number of targets in Fizeau mode\cite{Leisenring14, Conrad15, Conrad16, deKleer17, Hill13}. In particular, these previous observations were in ``lucky'' Fizeau mode, whereby the targets were imaged without active phase control and the ``lucky'' frames where the fringes were centered were selected out during the reduction phase. Short integration times were required in order to avoid phase smearing, which severely limited the sensitivity of Fizeau observations. There was also no way to know exactly which fringe had been centered, leading to a heavy loss of observable contrast due to the combination of phases. With the successful conclusion of HOSTS observations during the 2018A observing season\cite{Ertel18b}, the LBTI has begun the transition to commissioning controlled Fizeau imaging\cite{Spalding18}. Active, reliable phase control is critical to maintaining 0 OPD alignment of the Fizeau coherence envelopes for the duration of the observation. The grism-dispersed Fizeau fringes both immediately after the initial finding of the coherence envelope on the LMIRCam detector, as well as after minimization of the OPD, can be seen in Figure \ref{phase-lock-modes}. PHASECam was successfully used to track and stabilize fringes in a Fizeau observation of a bright star in May 2018 for up to a few minutes at a time, but fringe stability was seen to be interrupted by fringe jumps. 

NRM transforms large apertures into multi-element Fizeau interferometers by utilizing a pupil-plane mask to produce non-redundant baseline separations between apertures\cite{Leisenring12}. The power measured at certain spatial frequencies and position angles is associated with pairs of sub-apertures in the mask, allowing for the Fourier amplitudes and phases for each of the baselines to be measured. LMIRCam's 12-aperture pupil mask and the associated power spectrum for an observation of a bright calibrator, with points corresponding to both intra- and inter-aperture baselines, can be seen in Figure \ref{phase-lock-modes}. Previous NRM observations by the LBTI have been performed in single-aperture mode\cite{Sallum15}. However, similarly to the Fizeau mode, active phase control is required for maintaining alignment of the coherence envelopes for dual-aperture NRM observations\cite{Hinz16}. The first dual-aperture NRM observations were obtained using PHASECam in May 2018. However, also similarly to the controlled Fizeau observations, the fringe stability was occasionally interrupted by fringe jumps. 

The inefficiency of manual fringe jump correction, in combination with the cases of Fizeau and NRM interferometry, effectively illustrate the need for a more efficient, automated method of capturing and correcting fringe jumps. We therefore propose to combine the currently unused H band phase telemetry with the K band phase telemetry in order to implement multi-wavelength fringe tracking with PHASECam, which will allow automatic detection and correction of fringe jumps. We give a more detailed overview of PHASECam and its current approach to phase sensing and fringe jump capture in \S2, describe our new multi-wavelength approach and its implementation in \S3, present preliminary results and discussion thereof in \S4, and summarize and discuss future development in \S5.

\section{PHASECAM}

\subsection{Overview}\label{overview}

\begin{figure}[ht!]
\centering
\includegraphics[width=\textwidth]{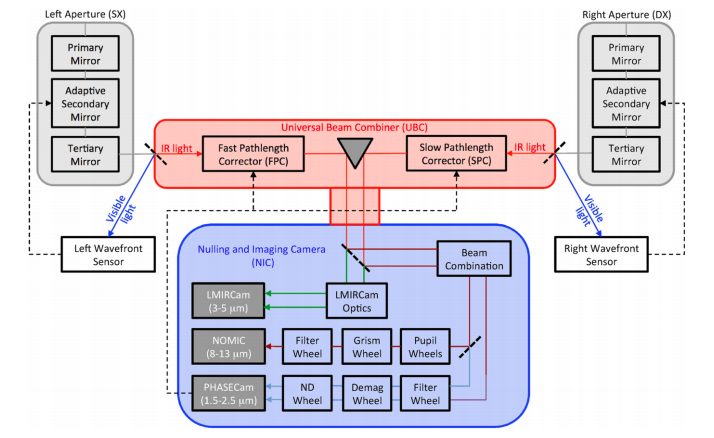}
\caption[angle=90]{System-level block diagram showing the optical light path through the telescope and the LBTI, including the beam combiner (red) and the NIC cryostat (blue). Visible light is reflected from the entrance window to the UBC to be used for wavefront sensing, while the infrared light passes into the cryogenic LBTI system. The beam combiner uses steerable mirrors to direct the light into the NIC cryostat, where the thermal near-infrared light (3-5 $\mu$m) is directed to LMIRCam for imaging and Fizeau interferometry, the mid-infrared (8-14 $\mu$m) light is directed to NOMIC for nulling interferometry, and the near-infrared (1.5-2.5 $\mu$m) light is directed to PHASECam for tip/tilt and phase sensing. PHASECam receives both outputs of the beam combiner, and sends tip/tilt and OPD corrections to the Fast and Slow pathlength correctors in the beam combiner (FPC/SPC). NOMIC receives only the nulled interferometric output, reflected by a short-pass dichroic. The three cameras are shown in dark gray, and the feedback loops between the wavefront sensors and the adaptive secondary mirrors as well as between the phase sensor and the pathlength correctors are shown by dashed lines. \textbf{This diagram is schematic only and does not show additional optics.} See [13].}
\label{system-block}
\end{figure}

PHASECam uses a fast-readout PICNIC detector which receives near-infrared light (1.5 - 2.5 $\mu$m) from both interferometric outputs when the system is arranged for either nulling or Fizeau interferometric imaging. A block optical path diagram of the telescope optics, LBTI's Universal Beam Combiner\cite{Hinz04} (UBC) and the Nulling and Imaging Camera\cite{Hinz08} (NIC) cryostat where PHASECam, NOMIC, and LMIRCam are housed can be seen in nulling mode in Figure \ref{system-block}\cite{Defrere16a}. PHASECam utilizes re-imaging optics to produce pupil images of each output beam, which are currently observed using standard H (1.65 $\mu$m) and K (2.2 $\mu$m) filters.

\begin{figure}[ht!]
\centering
\includegraphics[width=\textwidth]{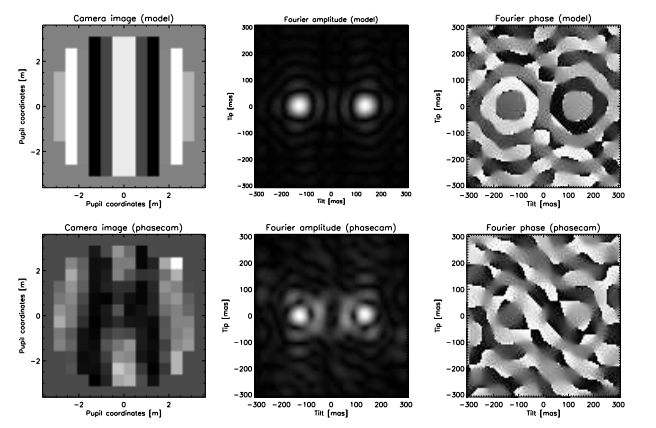}
\caption[angle=90]{LBTI's approach to phase sensing demonstrated for one interferometric output for a noise-free model (top) and on-sky K band data taken on March 14, 2017 (bottom). Pupil images of the interferometric output are imaged by PHASECam (left), and the Fourier transform (FT) of each image is used to perform tip/tilt and phase sensing. The position of the peak amplitude of the FT (middle) measures the differential tip/tilt, while the argument of the FT at that position (right) measures the phase. See [13].}
\label{pupil-images}
\end{figure}

The current approach to phase sensing and control uses the K band pupil images. When the images are well overlapped at the science wavelength of 10 microns, dispersion in the beamsplitter between 2 and 10 $\mu$m leads to a tilt difference of approximately three fringes across the pupil at 2 $\mu$m. This produces a signal in the Fourier plane which is well separated from the zero-frequency component, from which the differential tip/tilt and phase variations can be derived using a Fourier transform (FT)\cite{Defrere14}. The position of the peak amplitude of the FT measures the differential tip/tilt, and the argument of the FT at that position measures the differential phase.  Figure \ref{pupil-images}\cite{Defrere14} shows a K band pupil image with fringes, the amplitude of the FT, and the argument of the FT for a noise free model and for on-sky data. 

During closed loop operation the K band phase measurements are translated to OPD corrections at a rate of 1 kHz and send to the Fast and Slow Pathlength Correctors (FPC/SPC), piezo-electric transducers (PZTs) located in the UBC, which can also be seen in Figure \ref{system-block}. These measurements are also carried out on the H band output, which has been previously used to measure phase dispersion and water vapor variations between the two outputs\cite{Defrere16b}. However, as previously mentioned, the H band phase telemetry is not currently used for active control of the phase.

\subsection{Phase Pinning and Unwrapping}\label{pin-unwrap}

Due to the nature of the FT, the measured, or ``raw,'' phase is limited to a range of [-$\pi$, $\pi$]. The setpoint for the phase control law is unconstrained - that is, it is allowed to be any value, even if it is well outside the range of [-$\pi$, $\pi$]. Since the purpose of the phase control law is to change the pathlength value until the measured phase is equal to the setpoint, the measured phase needs to be modified for use in the control algorithm. In addition, we would like the maximum error that can be measured ($\pi$) to not be reduced by the choice of setpoint. 

To achieve both of these, we calculate a modified quantity called the ``pinned phase,'' which centers the measured phase values with respect to the setpoint and thus improves the Gaussianity of the noise statistics. While the range of the measured phase is always [-$\pi$, $\pi$], the range of the pinned phase is [setpoint-$\pi$, setpoint+$\pi$]. To calculate the pinned phase, if the setpoint is $\geq$0 and the difference between the setpoint and the measured phase is $>\pi$, 2$\pi$ is added iteratively to the measured phase until it is within $\pi$ of the setpoint. Conversely, if the setpoint is $<0$ and the difference between the setpoint and the measured phase is $<-\pi$, 2$\pi$ is iteratively subtracted from the measured phase until it is within $\pi$ of the setpoint.

\begin{figure}[ht!]
\centering
\includegraphics[width=\textwidth]{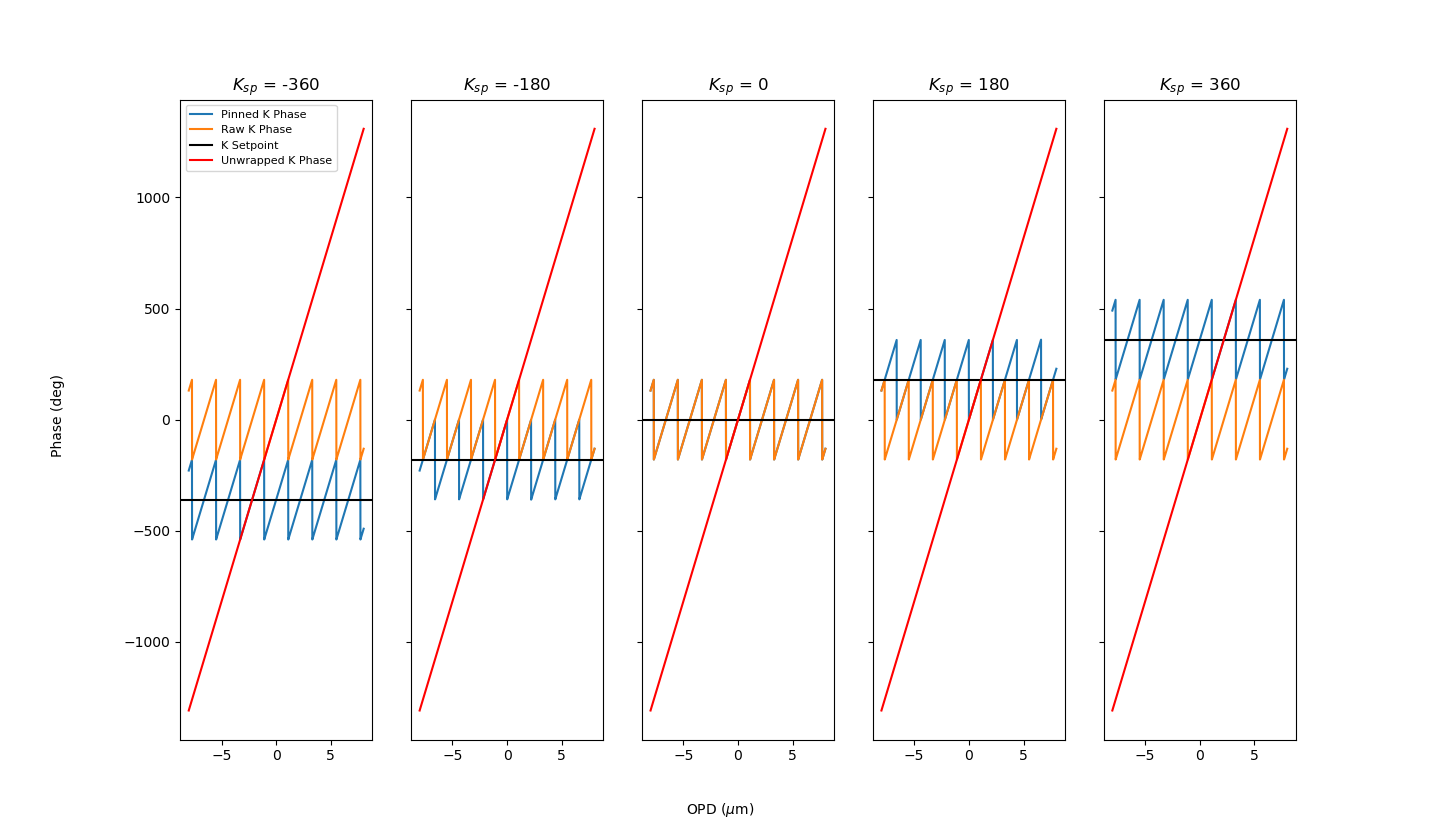}
\caption[angle=90]{The raw and pinned phases for a sample of positive and negative K band setpoints, for an ideal linear OPD scan from -8 to 8 $\mu$m, plotted here in phase space.}
\label{wrapped-phase}
\end{figure}

The raw and pinned phase values for a sample of setpoints can be seen for an ideal linear OPD scan from -8 to 8 $\mu$m in Figure \ref{wrapped-phase}. However, as can be seen, even when using the pinned phase, variations from the setpoint larger than $\pi$ are still not calculated correctly. Since OPD variations are often slowly varying, but of large amplitude, an improvement can be made to the dynamic range of the error by using the difference in the K band pinned phase between one time step and the next to ``unwrap'' it\cite{Colavita10}. If this difference is greater than $\pi$, or less than $-\pi$, it is assumed that the pinned phase crossed the $\pi$ boundary into a neighboring fringe and was subsequently wrapped. To correct this, $2\pi$ is added or subtracted from the pinned phase accordingly, giving the final unwrapped phase value for the given time step.

\subsection{Capturing Fringe Jumps}\label{fringe-jump}

The phase unwrapping algorithm is generally reliable at tracking phase variations. However, on-sky testing has revealed that even during 1 kHz closed-loop operation, phase variations occasionally occur which are large enough to cause a fringe jump.

\begin{figure}[ht!]
\centering
\includegraphics[width=\textwidth]{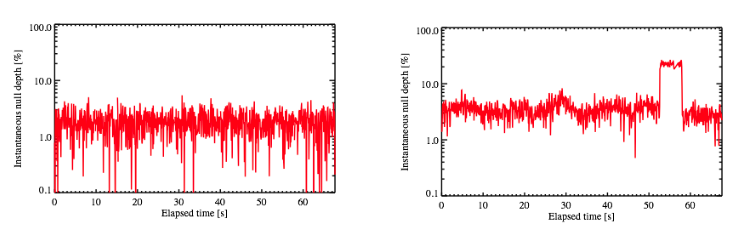}
\caption[angle=90]{Sample nulling sequences, with the jump away from null indicative of a fringe jump displayed on the right.}
\label{null-sequence}
\end{figure}

Previously, PHASECam detected and captured fringe jumps by tracking the envelope of interference, or the group delay. It utilized a metric known as the contrast gradient (CG) which tracked changes in the fringe contrast\cite{Defrere14}. LBTI's first stable, phase-tracked fringes were obtained in December 2013 using the CG. However, the precision of this method was limited to a closed-loop residual OPD of approximately 1 $\mu$m. Furthermore, the non-Gaussian phase distribution of this method complicated data reduction techniques. In 2015, PHASECam transitioned to phase delay tracking. With this method, fringe jumps are corrected manually by the observer. PHASECam's primary use has been for nulling interferometry, where a fringe jump presents as a distinctive departure from the typical null intensity, as seen in Figure \ref{null-sequence}. When a fringe jump is observed in the null telemetry, the observer adjusts the K band setpoint in increments of 2$\pi$ until the intensity returns to null.

This manual correction method works, and has been used for on-sky operation for several observing seasons. However, correction can take some time, both due to the amount of time it may take the observer to notice a fringe jump, and the fact that the observer may initially correct in the wrong direction. Furthermore, as previously described, for Fizeau interferometry and NRM, there is not a method such as observing the null telemetry for us to quickly detect fringe jumps. They can be seen if one closely examines the images of the fringes\cite{Spalding18}, but it is easily missed if one is not consistently watching. 

\section{A TWO-BAND APPROACH}

As previously mentioned, PHASECam simultaneously measures the phase telemetry in both H and K bands; however, it currently only makes use of the K band telemetry for phase control. We describe here a method which will make use of both bands of phase telemetry simultaneously in order to automatically determine whether the phase is in the central fringe or has jumped plus or minus one fringe relative to the phase setpoint. This equates to a full phase variation detection amplitude of $\sim$6$\pi$ in the K band/$\sim$8$\pi$ in H band, which adequately captures the typical range of phase variations seen by PHASECam. 

\begin{figure}[ht!]
\centering
\includegraphics[width=\textwidth]{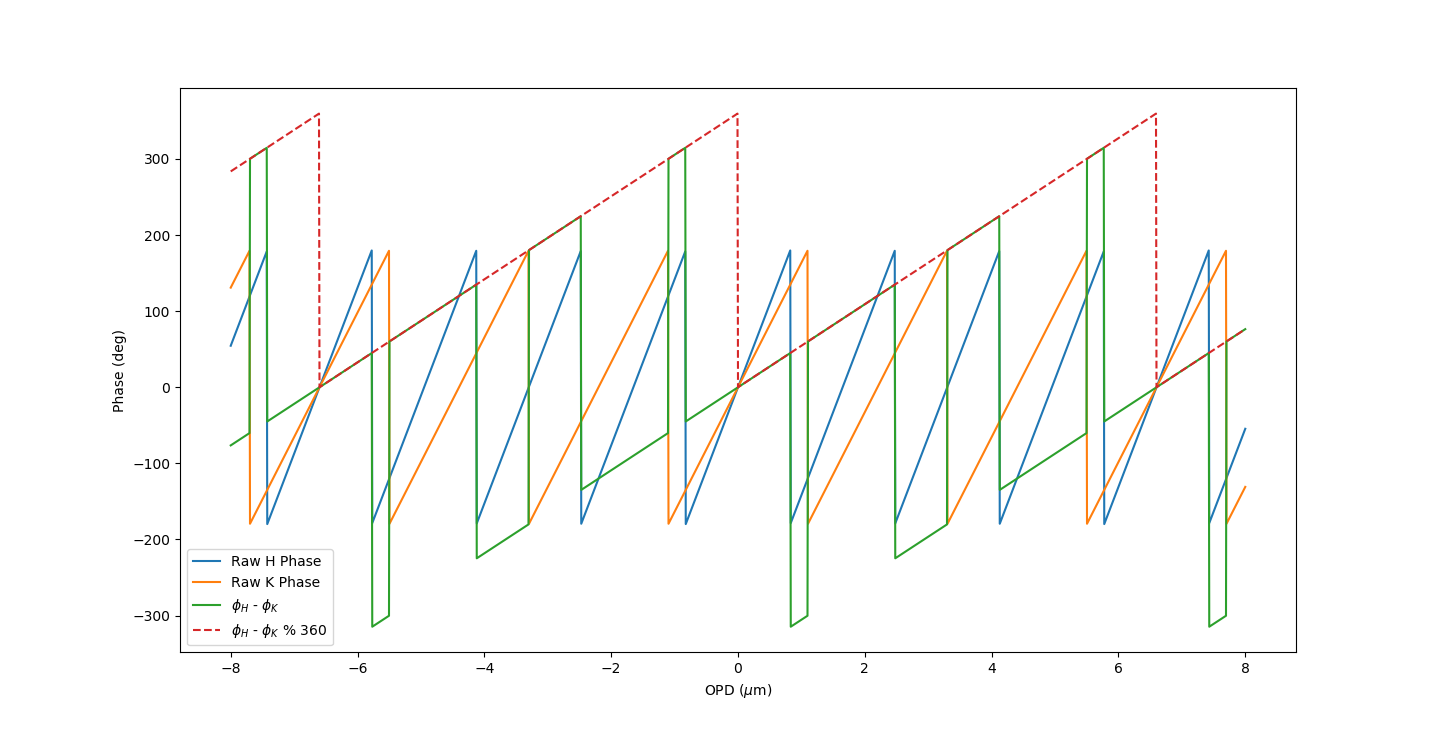}
\caption[angle=90]{Raw H and K band phases for a K band and H band setpoint of 0, with the simple difference and the diffmod metric overplotted. The multiple discontinuities in the simple difference can be seen, along with the single discontinuity in the diffmod at multiples of 1080. The diffmod, however, is single valued over the range from 0 to 1080. This is 3 fringes (6$\pi$) in K band and 4 fringes  (8$\pi$) in H band, which allows us to detect whether the phase has jumped plus or minus one fringe relative to the setpoint.}
\label{diffvsdiff}
\end{figure}

\subsection{The Difference-Modulo Metric}

The core of this two-band approach to fringe jump capture and correction is what we call the \textbf{difference-modulo metric}, henceforth referred to as the \textbf{diffmod}. The diffmod is mathematically described by

\begin{equation}
diffmod = (\phi_{H} - \phi_{K})\ \%\ 360.
\end{equation}

where $\phi_{H}$ and $\phi_{K}$ are the raw H and K phases, and \% represents the modulo operation. The version used here is the ``true'' modulo operation, typically seen in mathematics and also used in Python, which has been used for the development and initial testing of this method\footnote{See \S5.6: Binary Arithmetic Operations, here: \href{https://docs.python.org/2/reference/expressions.html}{https://docs.python.org/2/reference/expressions.html}}. The necessity of the modulo operation is demonstrated in Figure \ref{diffvsdiff} with the same linear OPD scan as in Figure \ref{wrapped-phase}. Simple differencing of the raw phases produces a complex result with multiple discontinuities due to the wrapping of the raw phases, which makes comparison difficult. The modulo, on the other hand, produces a smooth result that progresses from 0 to 360 from 0 phase out to the first common multiple of the observing wavelengths, which in the case of H and K band is 1080. It is discontinuous at multiples of the common multiple, but this single discontinuity is relatively simple to account for, as we will describe. 

Referenced to K band, the diffmod has a slope of 1/3: a fringe in K band phase space equates to 120 units in diffmod space, while a fringe in H band phase space equates to 90 units in diffmod space. We choose to reference to K band as while it is possible for the H band phase to pass into a neighboring fringe without the K band doing so - the discontinuous regions seen in Figure \ref{diffvsdiff} - these instances can be handled by the phase unwrapping algorithm. It is only when both the H and K band phases have consistently passed into a neighboring fringe that a true fringe jump can be said to have occurred, so, as we will describe, we look for the 120 unit diffmod change expected from K band rather than the 90 unit change expected from H band.

The details of the diffmod metric in practice can be seen graphically in Figure \ref{diffmodloop}. There are four steps:

\begin{figure}[ht!]
\centering
\includegraphics[width=.95\textwidth]{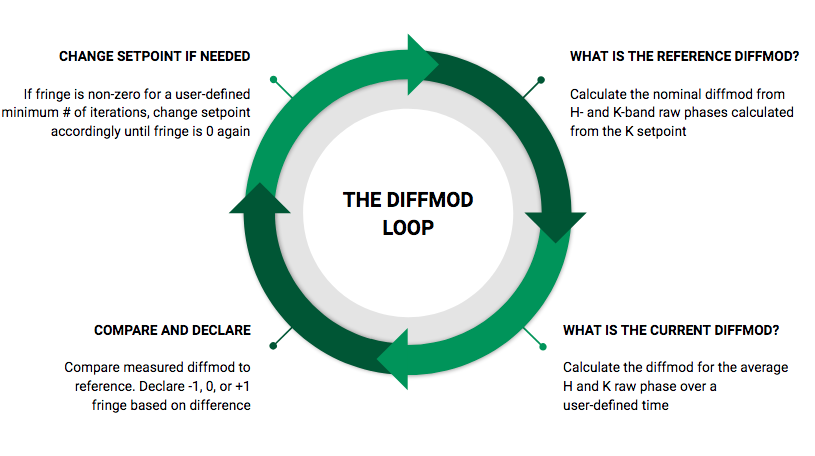}
\caption[angle=90]{The four core steps to the diffmod metric.}
\label{diffmodloop}
\end{figure}

\begin{enumerate}

\item \textbf{Calculate the reference diffmod.} Convert the current K band phase setpoint to an OPD using

\begin{equation}
OPD = K_{sp} \times \frac{\lambda_{K}}{360}.
\end{equation}

Then, convert this nominal OPD to raw H and K phases using

\begin{equation}
\phi_{\lambda,raw} = \left[\left(\left(OPD + \frac{\lambda}{2.0}\right)\ \times\ \frac{360}{\lambda}\right)\ \%\ 360\right] - 180.
\end{equation}

and calculate the diffmod of these two values. This is the nominal diffmod value to which the measured diffmod should average over time if there has not been a fringe jump, and it is recalculated when the K band setpoint changes.

\item \textbf{Measure the current diffmod.} Collect the measured raw H and K phases for a user-defined period of time. Average these phases, and then calculate the diffmod of the average. Averaging is necessary due to the noisy nature of the phase data; PHASECam has a residual OPD of approximately 300-400 nm rms\cite{Defrere16a}. Averaging prevents the metric from following rapid phase variations caused by noise which may cause it to declare a fringe jump where none occurred.

\item \textbf{Compare the measured and reference diffmods.} Calculate the difference between the measured and reference diffmods. If there has been a fringe jump, the diffmod over an averaging period will be approximately 120 greater or lesser than the reference diffmod, such that

\[fringe\ = \left\{
\begin{array}{ll}
      -1 & diffmod_{nom} - diffmod_{meas}\ \geq\ 120 \\
      0 & |diffmod_{nom} - diffmod_{meas}|\ <\ 120 \\
      1 & diffmod_{nom} - diffmod_{meas}\ \leq\ -120 \\
\end{array} 
\right. \]

A single instance of the phase jumping out of the central fringe, however, does not necessarily indicate a fringe jump. The phase may occasionally make transient jumps into a neighboring fringe. These typically do not persist, and are usually corrected by the phase unwrapping algorithm previously described
\begin{figure}[H]
\centering
\includegraphics[width=.8\textwidth, trim=2.5cm 0cm 2.5cm 0cm, clip=true, keepaspectratio]{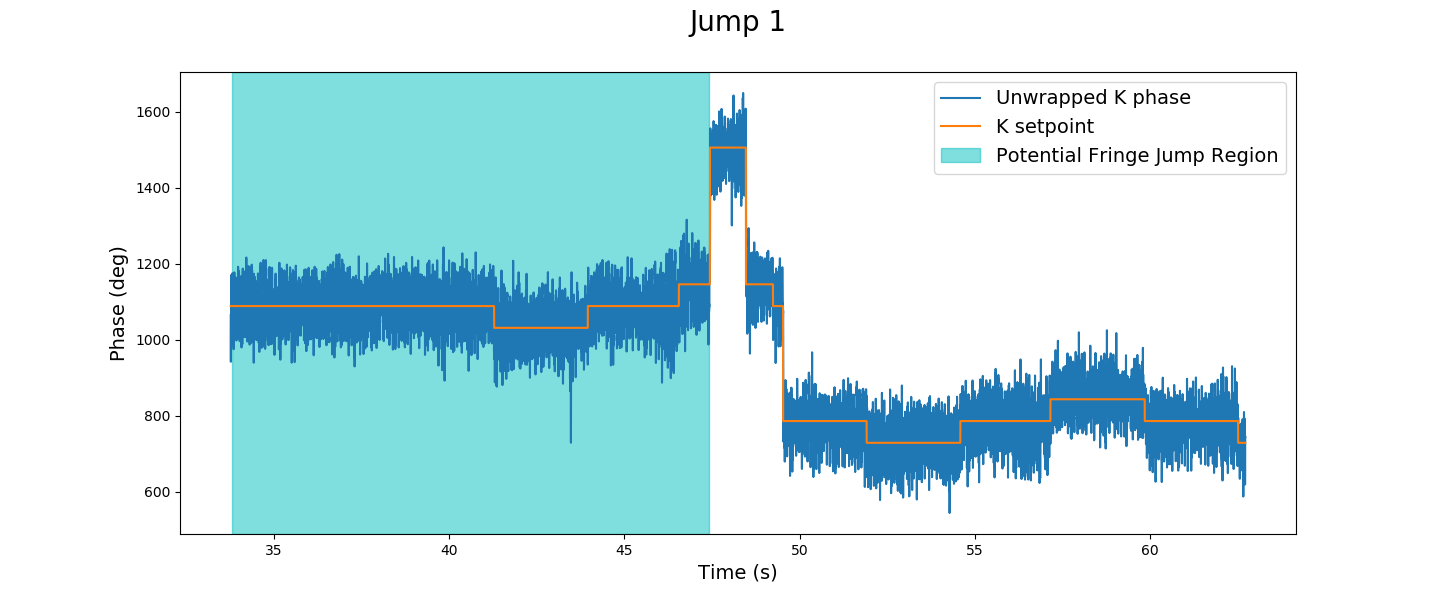}
\includegraphics[width=.8\textwidth, trim=2.5cm 0cm 2.5cm 0cm, clip=true, keepaspectratio]{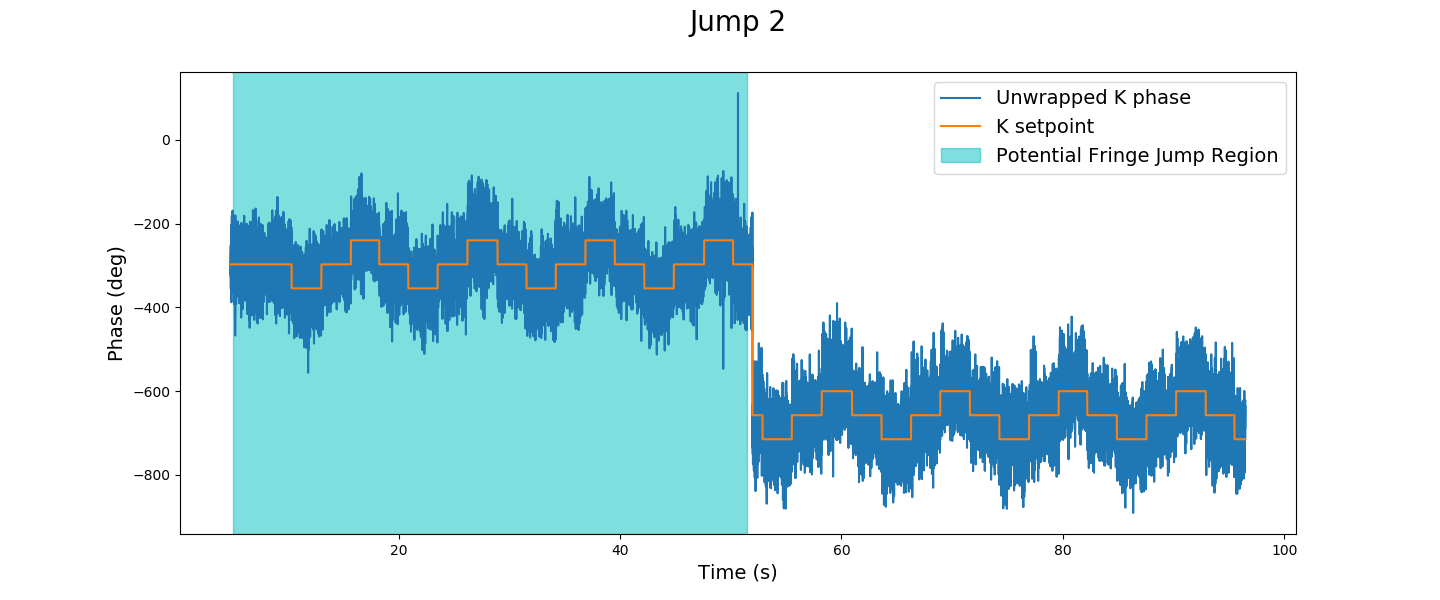}
\includegraphics[width=.8\textwidth, trim=2.5cm 0cm 2.5cm 0cm, clip=true, keepaspectratio]{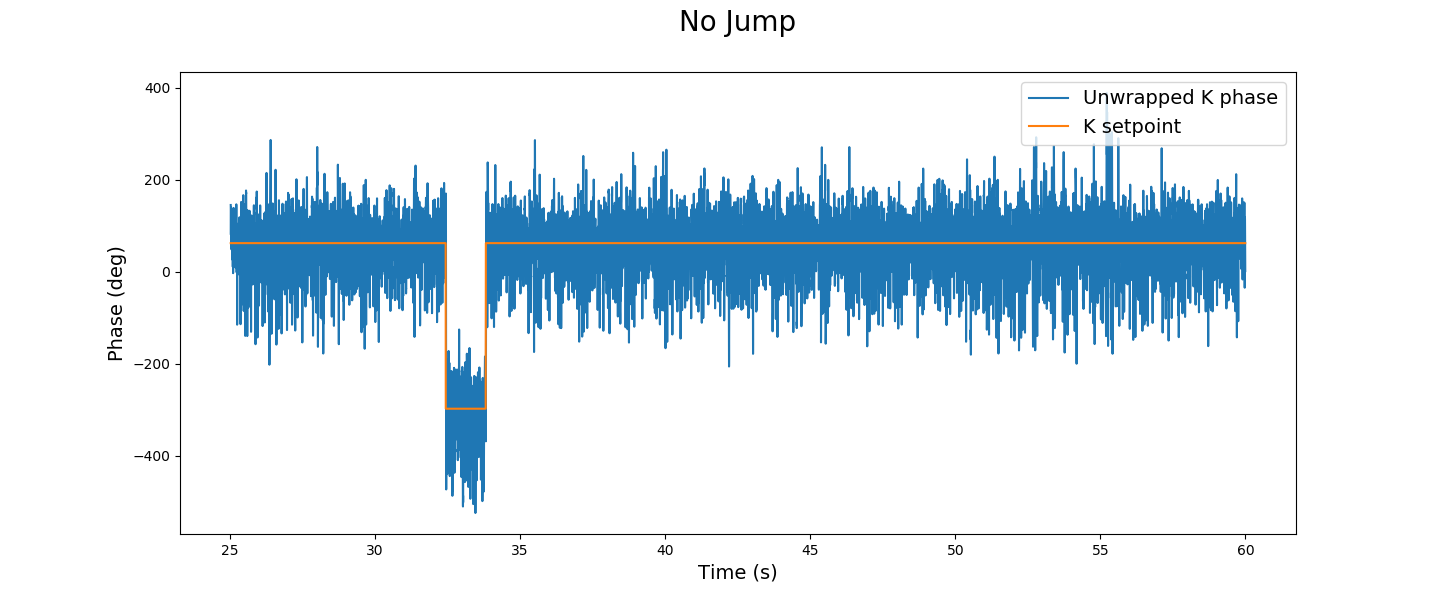}
\caption[angle=90]{Unwrapped K phase telemetry sequences, including two with known fringe jumps and one without. The regions of time over which the fringe jumps could have occurred are shaded. The small changes in setpoint seen in the Jump 1 and Jump 2 telemetry are from automated setpoint dithering/phase modulation during nulling observations, while the large setpoint changes in these sequences indicate manual correction of fringe jumps by the observer. The large setpoint change seen in the No Jump telemetry is another instance of setpoint dithering/phase modulation, instigated manually by the observer rather than the automated observing script.}
\label{unwrapped-telem}
\end{figure}in \S\ref{pin-unwrap}. It is only when the jump persists over a longer period of time that a fringe jump can be said to have occurred. We therefore require the loop outlined in Figure \ref{diffmodloop} to declare a non-zero fringe a minimum user-defined number of averaging periods in a row before truly declaring a fringe jump and activating the final step. This count is reset to 0 if the fringe value returns to 0 at any point in time.

\item \textbf{Change the setpoint.} If, as described in the previous step, the declaration of a non-zero fringe persists for a minimum number of averaging periods, a fringe jump is declared and this intervention step is activated, and the K band phase setpoint is changed by $\pm2\pi$ for a $\pm$1 fringe jump, respectively. The jump count described in the previous step is reset to 0 at this point. This four step process repeats until the fringe has reliably returned to 0, at which point it returns to the monitoring of the first three steps.

\end{enumerate}

In order to demonstrate the ability of the diffmod metric to determine relative fringe values and declare the presence of a fringe jump - the first three steps of the diffmod loop - we have analyzed three previously collected phase telemetry sequences. We are unfortunately unable to present data demonstrating the full loop including active setpoint adjustment. The diffmod metric was originally implemented and tested in a different form, which underwent several iterations over the course of the LBTI observing run from 22 May 2018 - 01 June 2018. The metric did not reach the form described here until after the conclusion of said run, which was the last for the LBTI of the 2018A observing season. We have instead selected segments of telemetry sequences from previous observing runs which were noted to contain multiple fringe jumps, as well as a representative sequence from the late May run which does not contain a fringe jump. The first two sequences, henceforth known as Jump 1 and Jump 2, are from a HOSTS science observation of Vega on UT 03 March 2018 and are both known to contain a fringe jump. The third sequence, henceforth known as No Jump, is from a HOSTS calibrator observation of HD199101 taken on UT 25 May 2018, and does not contain a fringe jump. The unwrapped phase telemetry for these three sequences can be seen in Figure \ref{unwrapped-telem}. 

For Jump 1, the fringe jump occurred somewhere between 34 and 48 seconds after UT 11:29. The setpoint was initially corrected in the wrong direction by the observer, as can be seen from the subsequent +2$\pi$ and then -2$\pi$ setpoint changes. For Jump 2, the fringe jump occurred somewhere between 5 and 51.5 seconds after UT 11:32, and the setpoint was successfully corrected by the observer on the first attempt. The small changes in K band setpoint seen in both the Jump 1 and Jump 2 telemetry demonstrate setpoint dithering. The No Jump telemetry sequence has no setpoint dithering as it was taken after the conclusion of observations but while PHASECam was still locked on the target. It does contain a brief setpoint adjustment, used in order to check the alignment with the central null.

\section{RESULTS AND DISCUSSION}

\begin{figure}
\centering
 \includegraphics[width=.75\textwidth, height=.3\textheight, trim=2.5cm 0cm 2.5cm 0cm, clip=true]{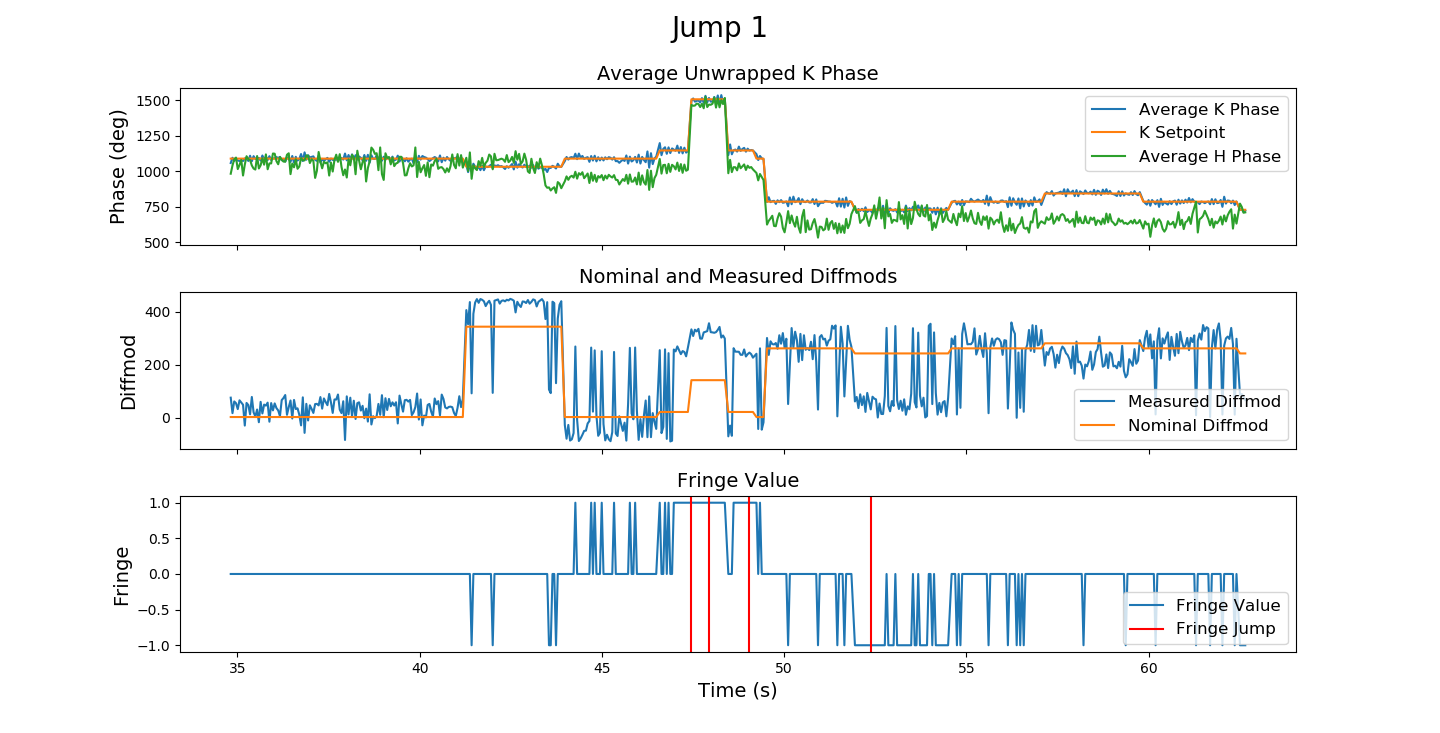}
\includegraphics[width=.75\textwidth, height=.3\textheight, trim=2.5cm 0cm 2.5cm 0cm, clip=true]{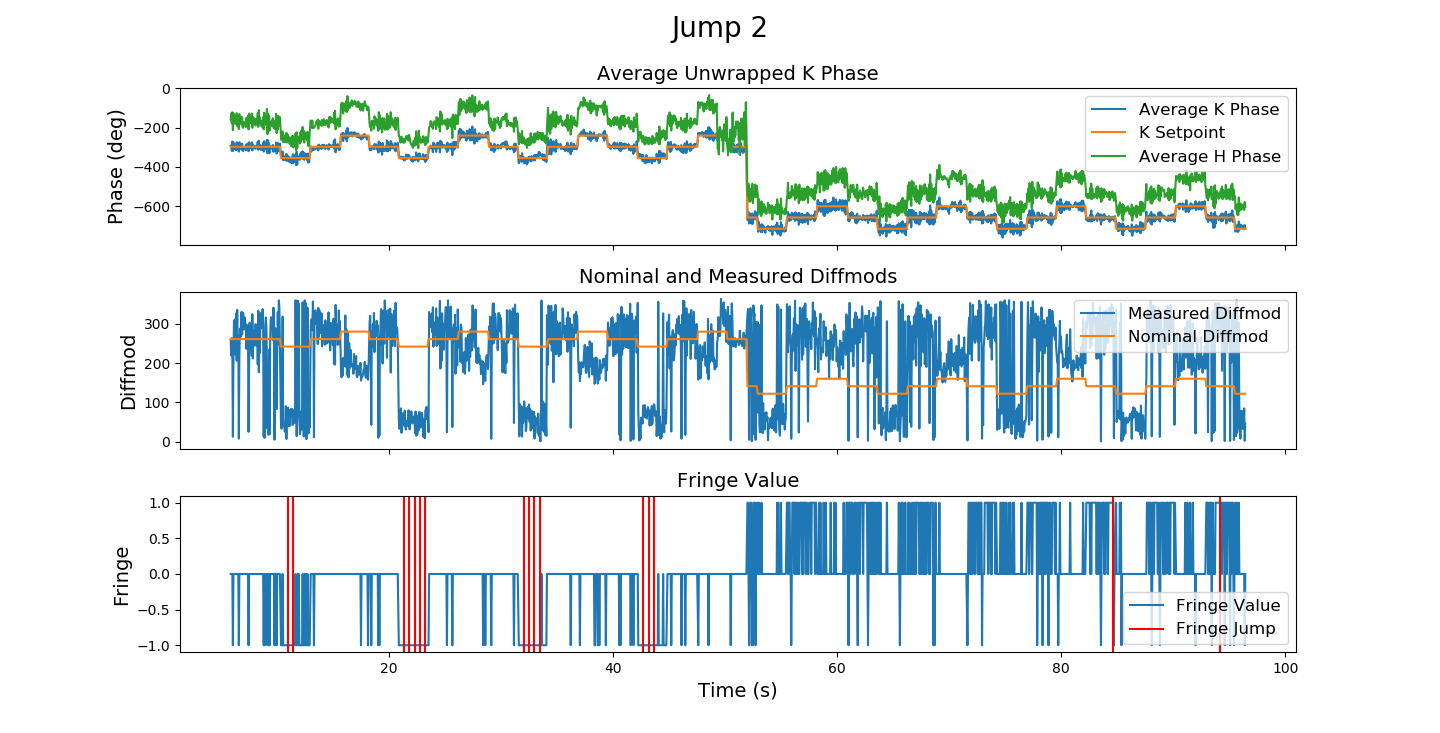}
\includegraphics[width=.75\textwidth, height=.3\textheight, trim=2.5cm 0cm 2.5cm 0cm, clip=true]{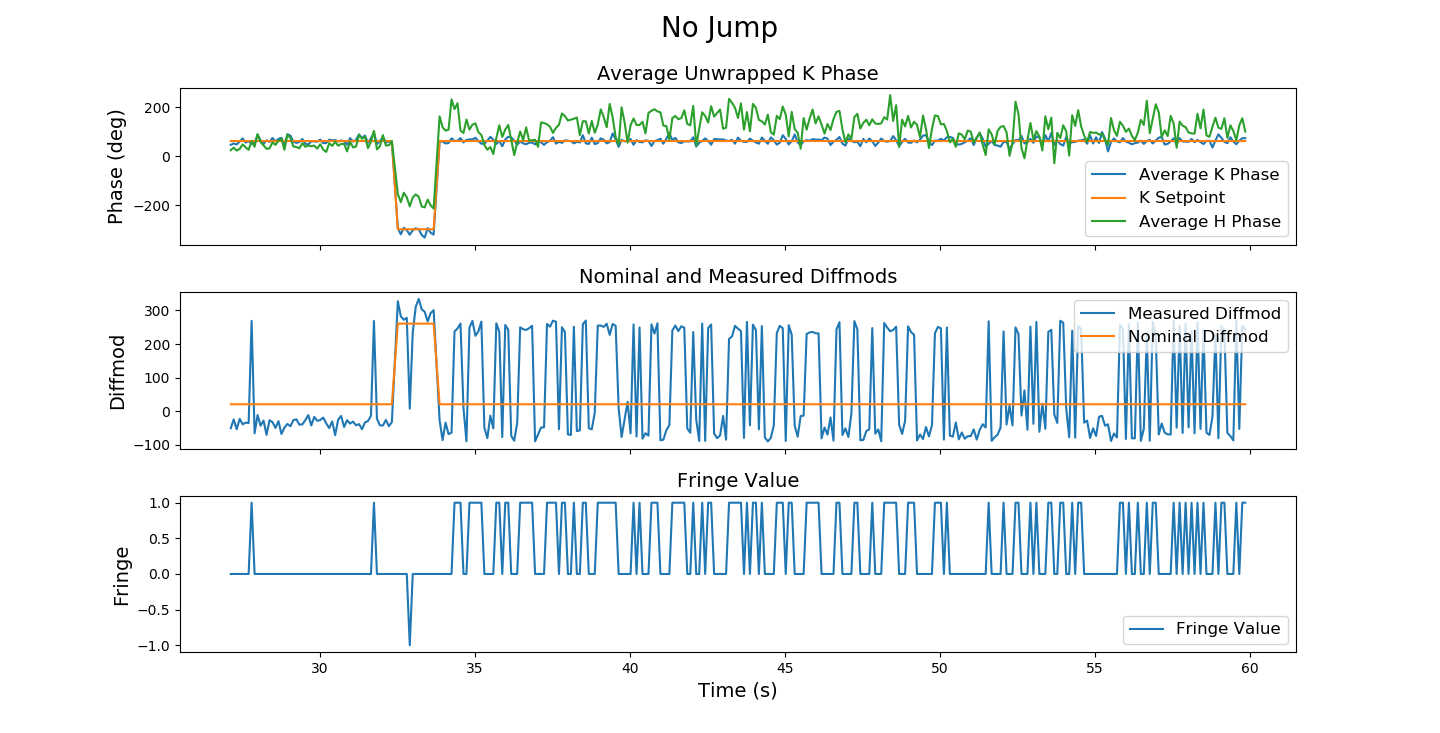}
  \caption[angle=90]{Results of the diffmod analysis for the three phase telemetry sequences presented in Figure \ref{unwrapped-telem}. For each sequence, the averaged H and K unwrapped phases (top), the nominal and measured diffmods (middle), and the fringe value for each averaging period (bottom) are shown. Instances of a fringe jump as determined by the diffmod loop are marked in red on the fringe value plot. The averaging period is 0.05 s, and the minimum number of non-zero fringe declarations is 10.}
\label{results}
\end{figure}

The results of our diffmod analysis of the three phase telemetry sequences can be seen in Figure \ref{results}. The averaging period for this analysis was approximately 0.05 seconds, with the minimum required number of declarations of a non-zero fringe value set to 10, for a jump detection period of approximately 0.5 seconds. For each sequence, the averaged H and K unwrapped phases, the reference and measured diffmod values, and the declared fringe value for each averaging period are shown, with the location of fringe jumps marked on the fringe jump plot. We describe the results for each sequence individually.

\subsection{Jump 1}

The diffmod loop successfully identifies Jump 1. The fringe value for Jump 1 can be seen to begin oscillating rapidly between the 0 and +1 fringes at approximately 43.5 s after UT 11:29. It transitions fully to the +1 fringe at 47 s, which is also the point at which the difference between the reference and measured diffmods can be seen to reliably increase significantly. 0.5 s later, a fringe jump is declared, which coincides with the observer's initial incorrect change of setpoint. The fringe value remains at +1; it is at +2 referenced to the correct setpoint, but still at +1 referenced to the new, incorrect setpoint. A fringe jump is declared again another 0.5 s later: the observer changes the setpoint back to the original value. A fringe jump is declared a third time, and the observer changes the setpoint again, this time in the correct direction; the fringe value finally reliably returns to 0. 

However, at approximately 52 s, the fringe value can be seen to jump to -1, and a fringe jump is declared again at 52.5 s. The fringe value returns to 0 at approximately 55 s. According to the unwrapped phase data, there is no fringe jump at this point, as no setpoint change occurs except for the small changes from setpoint dithering. In fact, a closer look at the data reveals this portion of the telemetry sequence to coincide with the lower point in the setpoint dither scheme. This begins to repeat at the very end of the displayed telemetry sequence, where the setpoint returns to this dither point again. The measured diffmod decreases much further than is to be expected during this setpoint dither. We discuss this phenomena in more detail with Jump 2.

\subsection{Jump 2}

\begin{figure}[ht!]
\centering
\includegraphics[width=\textwidth, keepaspectratio]{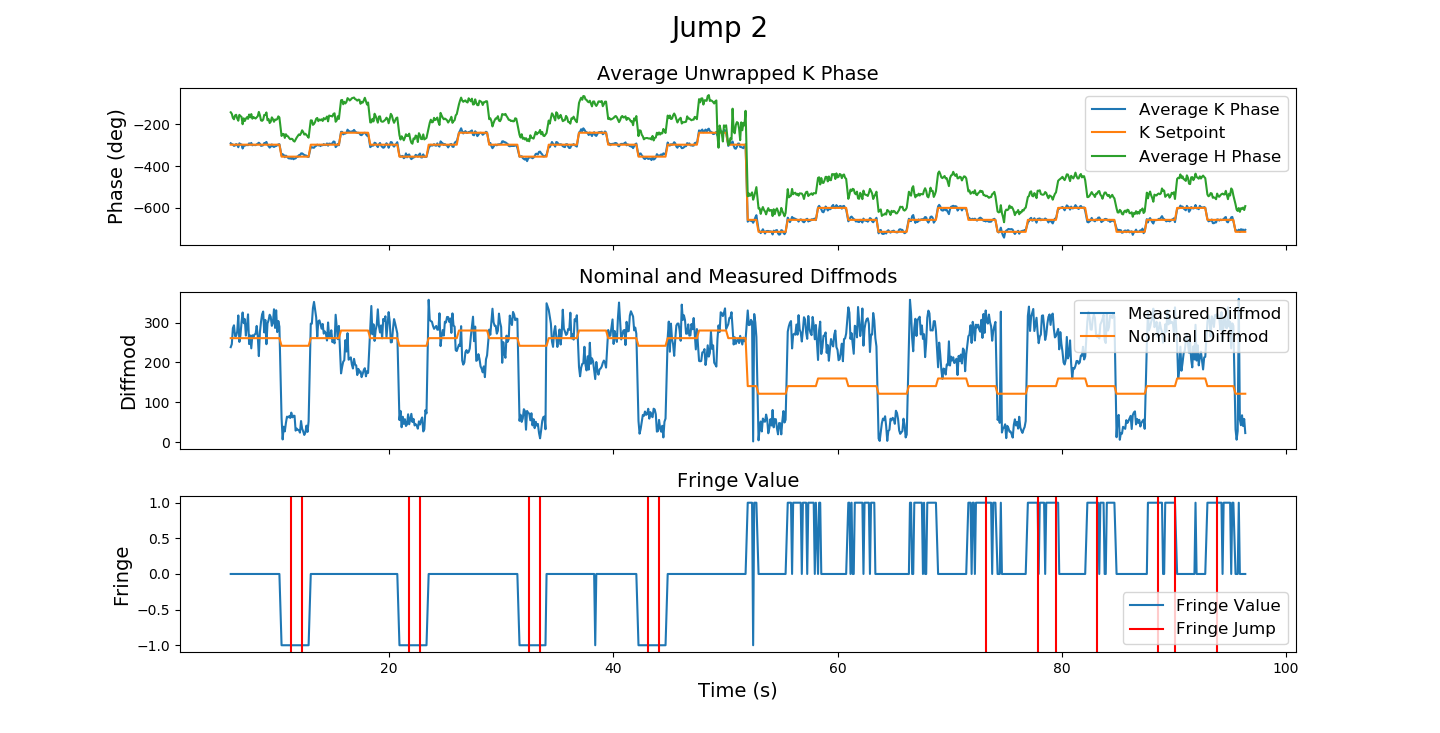}
\caption[angle=90]{Results of the diffmod analysis for Jump 2 with an averaging period of approximately 0.1 seconds. All fringe jumps prior to the setpoint change directly coincide with the lower point in the setpoint dither scheme. All fringe jumps after the setpoint change coincide with the central point of the dither scheme. }
\label{jump2-100}
\end{figure}

The diffmod loop does not successfully identify Jump 2. The fringe value can be seen to oscillate frequently between the -1 fringe and the 0 fringe prior to the setpoint change, and then between the +1 fringe and the 0 fringe after said change. However, the fringe value for several seconds prior to the setpoint change is predominantly 0. There are four points prior to the setpoint change at which the fringe value jumps to -1 for a short period of time and a fringe jump is declared. However, closer inspection reveals all four of these segments to coincide with the lower setpoint dither point, as previously seen with Jump 1. Fewer fringe jumps are declared following the setpoint change, but those that are declared coincide with the central dither point. If the averaging period for Jump 2 is increased to approximately 0.1 seconds, this phenomena becomes even clearer, as can be seen in Figure \ref{jump2-100}. 

Looking at the diffmod data, the measured diffmod for the lower dither segments decreases much further than expected, as also seen with the Jump 1 telemetry. In fact, the measured diffmod seems to not follow the expected pattern in several respects. Prior to the setpoint change, the measured diffmod decreases at the upper dither point when it is expected to increase. It does the same following the setpoint change. Additionally, the measured diffmod remains at approximately the same set of values as prior to the setpoint change following it, rather than decreasing in proportion with the change. 

\subsection{No Jump}

The diffmod loop successfully does not declare a fringe jump for the No Jump telemetry sequence. The fringe value remains reliably 0 before and during the setpoint change, as expected.

Following the setpoint change, however, the fringe value can be seen to begin to oscillate frequently between the 0 and +1 fringes. Looking at the unwrapped H phase telemetry, it can be seen that the telemetry is much noisier following the setpoint change. In this case, the oscillations do not last long enough for a fringe jump to be declared, but the issue is under investigation. It is not believed to have a physical cause: the H band unwrapping algorithm is not yet finalized.  

\subsection{Discussion}

Our diffmod analysis of the three phase telemetry sequences successfully identified one of the two known fringe jumps, and successfully did not identify a fringe jump in the phase telemetry without one. 

In both the sequences where it successfully and did not successfully identify the known fringe jump, the diffmod loop mis-declared further fringe jumps at locations in the phase telemetry coincident with the setpoint dithering away from the central setpoint value. This is at least partially due to the fact that the central K band setpoint is often at values close to the diffmod discontinuity. The combination of setpoint dithering and noise may cause the phase to consistently cross the discontinuity and be wrapped to a value that appears to indicate a fringe jump, though one did not occur.

Currently, to at least partially mitigate this issue, we have implemented a check on the nominal and measured diffmods which says if both are within approximately $\frac{3}{4}$ of a K band fringe - 90 diffmod units - of the discontinuity but on opposite sides of it, then a wrap has most likely occurred. The check reassigns the measured diffmod value to a value greater than 360 or less than 0 accordingly, to allow a meaningful comparison to be made. For example if the nominal diffmod is 20 and the measured diffmod is 350, the measured diffmod is reassigned to -10, since it is far more likely to still be in the central fringe than to have jumped over two fringes, which is what the diffmod loop would declare. Another solution we are investigating is dynamically shifting the diffmod so that the nominal point is always centered in the diffmod range at 180, so that the diffmod never crosses the discontinuity in the middle of a fringe. 

However, these issues with mis-declarations of fringe jumps may also be indicative of a different issue. In our implementation of the diffmod metric, by calculating both nominal raw phase values from the K band setpoint, we have implicitly assumed that the H band setpoint is the setpoint calculated from the ideal (4/3) proportion between the H and K phases. In practice, this is not the case. The H and K band setpoints have a differential offset, the predominant cause of which is the placement of the software apertures over which phase sensing calculations are done. Dispersion in the optical components of PHASECam and dispersion from water vapor also contribute. There is a calculation of this offset value in the PHASECam software, but its implementation with regards to fringe jump capture is still being revised. Implementing this, alongside finalizing the H band phase unwrapping algorithm as previously mentioned should significantly increase the accuracy of the diffmod loop. Correct determination of the H band setpoint will also allow us to make the transition from using the raw H and K band phases to the pinned phases. As previously described, this will center the measured phase values so that the maximum measurable error is not reduced by the choice of setpoint, increasing the Gaussianity of the noise statistics which should further increase the accuracy of the diffmod.  

With regards to noise statistics, another aspect of the diffmod loop which showed itself to be important during the analysis was the averaging period. While PHASECam is quite successful at removing the effects of instrument flexure ($\ll$ Hz) and the atmosphere ($\sim$10 Hz), it is less successful at removing the effects of low frequencey telescope resonances at 12-18 Hz and higher-frequency instrument vibrations at 100-150 Hz. All of these contribute significantly to the previously described 300-400 nm rms residual OPD seen in the PHASECam telemetry. Averaging can reduce the noise effects, but requires a delicate balance. Too short of an averaging period leaves the diffmod loop prone to declaring fringe jumps on transient OPD variations, and too long begins to smooth out the telemetry features we are interested in. Though 0.05 seconds was used as the averaging time for the primary analysis, Figure \ref{jump2-100} demonstrated how much of an effect an increased averaging time could have on the data. The optimum averaging time seems to be somewhere between 0.05 and 0.1 s, but whether that value is static or something that need to be determined dynamically remains to be seen. 

\section{SUMMARY AND FUTURE WORK}

We have developed the difference-modulo, i.e., ``diffmod'' metric in order to transition LBTI's fringe tracker, PHASECam, from manual to automatic fringe jump capture and correction. The diffmod utilizes both the H and K band phase telemetry simultaneously, versus the previous method which only utilized the K band telemetry. It compares the difference of the average H and K band raw phases modulo 360 to the nominal difference based upon the phase setpoint modulo 360, in order to declare whether the phase value is in the central (0), +1, or -1 fringe relative to the setpoint. The diffmod is valid for any two observing wavelengths, and is single-valued from 0 phase to the first common multiple of said wavelengths.

Testing of the diffmod metric showed it to be promising, but not yet completely reliable at identifying fringe jumps. The primary reasons for this seem to be (1) The discontinuity in the diffmod metric at multiples of 1080, which typical values of the K band setpoint are often near and which can cause fringe jumps to be declared if the diffmod wraps. (2) Residual OPD from the atmosphere, telescope resonances, and instrument vibrations causing noise in the diffmod values. (3) The differential setpoint between H and K bands, caused by software aperture placement, dispersion in optical components, and dispersion from water vapor. The current calculation of this value in the PHASECam software, as well as the H band phase unwrapping algorithm, are still being revised. 

We will be working to mitigate these issues in the coming months, by implementing dynamic centering of the diffmod, iterating the averaging period to determine the ideal value which reduces the effects of the OPD residuals while not smoothing out relevant telemetry features, and completing the revision of the PHASECam code with regards to H band phase unwrapping and setpoint determination. We also plan to transition the diffmod to use of the pinned phases rather than the raw phases in order to increase the dynamic range of the error and improve its response to noise, but this requires accurate knowledge of the H band setpoint. 

NIC, including PHASECam, is currently removed from the LBT for annual testing and upgrades. PHASECam itself will be undergoing several upgrades to its optics and electronics to improve its limiting magnitude. We aim to begin on-sky real-time testing of the diffmod metric in the 2018B LBTI observing season, when PHASECam is returned to the LBT, in tandem with further development of the controlled Fizeau and NRM modes.

\acknowledgements
LBTI is funded by a NASA grant in support of the
Exoplanet Exploration Program (NSF 0705296).
The LBT is an international collaboration among institutions in the United States, Italy and Germany. LBT Corporation partners are: The University of Arizona on behalf of the Arizona university system; Istituto Nazionale di Astrofisica, Italy; LBT
Beteiligungsgesellschaft, Germany, representing the Max-Planck Society, the Astrophysical Institute Potsdam, and Heidelberg University; The Ohio State University, and The Research Corporation, on behalf of The University of Notre Dame, University of
Minnesota and University of Virginia.

\bibliography{report}
\bibliographystyle{spiebib}

\end{document}